\documentclass[aip, amsmath,amssymb,preprint]{revtex4-1}
\usepackage{braket}
\usepackage{graphicx}
\usepackage{dcolumn}
\usepackage{bm}
\usepackage{subfig}
\usepackage[utf8]{inputenc}
\usepackage[T1]{fontenc}
\usepackage{amsmath}
\usepackage{xcolor}
\usepackage{epstopdf}

\usepackage{booktabs}

\begin{document}
\title{Local self-interaction correction method with a simple scaling factor}
\author{Selim Romero}
\affiliation{Department of Physics, University of Texas at El Paso, El Paso, Texas 79968, USA}
\affiliation{Computational Science Program, University of Texas at El Paso, \\ El Paso, Texas 79968, USA}
\author{Yoh Yamamoto}
\affiliation{Department of Physics, University of Texas at El Paso, El Paso, Texas 79968, USA}
\author{Tunna Baruah}
\affiliation{Department of Physics, University of Texas at El Paso, El Paso, Texas 79968, USA}
\affiliation{Computational Science Program, University of Texas at El Paso, \\ El Paso, Texas 79968, USA}
\author{Rajendra R. Zope}
\email{rzope@utep.edu}
\affiliation{Department of Physics, University of Texas at El Paso, El Paso, Texas 79968, USA}
\affiliation{Computational Science Program, University of Texas at El Paso, \\ El Paso, Texas 79968, USA}

\date{\today}

\begin{abstract}
A recently proposed local self-interaction correction (LSIC) method [Zope \textit{et al.} J. Chem. Phys., 2019,{\bf 151}, 214108]
when applied to the simplest local density approximation provides significant improvement over standard Perdew-Zunger SIC (PZSIC) for 
both equilibrium properties
such as total or atomization energies  as well as properties involving stretched bond such as barrier heights.
The method uses an iso-orbital indicator to identify the single-electron regions. To demonstrate the LSIC method, 
Zope \textit{et al.} used the ratio $z_\sigma$ of von Weizs\"acker $\tau_\sigma^W$ and total kinetic energy 
densities $\tau_\sigma$, ($z_\sigma = \tau_\sigma^W/\tau_\sigma$) as a scaling factor to scale  the self-interaction
correction. The present work further explores the LSIC method using a simpler scaling factor as a ratio  
of orbital and spin densities in place of the ratio of kinetic energy densities. 
We compute a wide array of both,
equilibrium and non-equilibrium properties using the LSIC and orbital scaling methods using this simple scaling factor
and compare them with previously reported results. Our study shows that the present results with simple 
scaling factor are comparable to those obtained by LSIC($z_\sigma$) for most properties but have slightly
larger errors. We furthermore study the binding energies of 
small water clusters using both the scaling factors. Our results show that LSIC with $z_{\sigma}$
has limitation in predicting the binding energies of weakly bonded system due to the inability of
$z_{\sigma}$ to distinguish weakly bonded region from slowly varying density region. LSIC
when used with density ratio as a scaling factor, on the other hand, provides  good description of water
cluster binding  energies, thus highlighting
the appropriate choice of iso-orbital indicator.
\end{abstract}

\maketitle

\section{\label{sec:introduction} Introduction} 
Kohn-Sham (KS) formulation of density functional theory (DFT)\cite{PhysRev.140.A1133,jones2015density,RevModPhys.61.689}
is widely used 
to study electronic structures of atoms, molecules, and solids because of its 
low computational 
cost and availability of easy to use software packages. 
The practical application of DFT requires an approximation to the exchange-correlation (XC) functional. 
The simplest form of the XC functional is the local spin density approximation (LSDA)\cite{PhysRev.140.A1133,PhysRevB.23.5048}
which belongs to the lowest rung of ladder of  the XC functionals\cite{doi:10.1063/1.1390175}.
The higher rungs of the ladder contains more complex and more accurate functionals- 
generalized gradient approximation (GGA), meta-GGA, hybrid, and functionals that include the 
virtual orbitals.
Practically all efforts in the functional design have been 
focused on improving the energetics or equilibrium properties such as atomization energies, bond distances, etc.
The majority of the density functional approximations suffer from self-interaction errors (SIE) though the magnitude of error 
can vary from one class of functionals to another or from one parameterization to another in a given class of functional.
The SIE occurs as a result 
of incomplete cancellation of self-Coulomb energy by the self-exchange energy of the approximate XC functional.

   Many failures of density functional approximations (DFAs) have been attributed to the SIE. 
   The SIE causes the potential to decay asymptotically as $-exp(-r)$ instead of the 
   correct $-1/r$ decay for finite neutral systems.  As a result the DFAs produce errors 
   such as too shallow eigenvalues of valence orbitals, inaccurate chemical reaction 
   barriers, electron delocalization errors, incorrect charges on dissociated fragments,
   incorrect binding energies for anions, etc.\cite{PhysRevB.23.5048,doi:10.1063/1.3021474,doi:10.1063/1.4829642,doi:10.1063/1.1630017}
   The $-1/r$ asymptotic behavior is also important for the computation of electronic properties that are sensitive to virtual 
   orbitals and long-range density such as excited states for example.

A number of approaches to remove the SIEs have been proposed.\cite{lindgren1971statistical,PhysRevA.15.2135,perdew1982density, lundin2001novel,doi:10.1063/1.2403848,gidopoulos2012constraining,doi:10.1002/jcc.10279,doi:10.1063/1.2204599,doi:10.1063/1.5129533,
doi:10.1063/1.2403848, 
PhysRevB.76.033102, 
doi:10.1063/1.4866996}
Early approaches\cite{lindgren1971statistical,PhysRevA.15.2135} 
 used  orbitalwise schemes to eliminate the SIE but used functionals related to Slater's X$\alpha$ method \cite{slater1951simplification}.
 The most widely used approach to remove SIE is the one proposed by  Perdew and Zunger (PZ)\cite{PhysRevB.23.5048}.
 Their approach is commonly referred to as PZ self-interaction correction (PZSIC) where the 
 one-electron SIE due to both exchange and correlation are removed from a DFA calculation on
 an orbital by orbital basis. 
PZSIC provides the exact cancellation for 
one- and two-electron self-interaction (SI), but not necessarily for many-electron SI\cite{doi:10.1063/1.2566637}.
It  has been applied to study properties of atoms, molecules, clusters, and 
    solids.\cite{ doi:10.1063/1.481421, doi:10.1063/1.1327269, doi:10.1063/1.1370527, harbola1996theoretical, doi:10.1063/1.1468640, doi:10.1021/jp014184v,PhysRevA.55.1765,doi:10.1080/00268970110111788, Polo2003, doi:10.1063/1.1630017, B311840A,doi:10.1063/1.1794633, doi:10.1063/1.1897378, doi:10.1063/1.2176608, zope1999atomic,  zope2000momentum,fois1993self,doi:10.1063/1.2204599,doi:10.1002/jcc.10279,PhysRevA.45.101, PhysRevA.46.5453,lundin2001novel, PhysRevA.47.165,doi:10.1021/acs.jctc.6b00347,csonka1998inclusion,petit2014phase,kummel2008orbital,schmidt2014one,kao2017role,doi:10.1002/jcc.25586,jonsson2007accurate,rieger1995self,temmerman1999implementation,daene2009self,szotek1991self,messud2008time,messud2008improved,doi:10.1063/1.1926277,korzdorfer2008electrical,korzdorfer2008self,ciofini2005self,PhysRevA.50.2191,
    doi:10.1063/1.5125205,C9CP06106A,doi:10.1002/jcc.25767,doi:10.1021/acs.jctc.8b00344,doi:10.1063/1.4947042,schwalbe2019pyflosic,
    doi:10.1063/1.4996498, doi:10.1063/1.5050809, doi:10.1021/acs.jpca.8b09940,Jackson_2019,Sharkas11283}

The PZSIC is an orbital dependent theory and when used with the KS orbitals results in size-extensivity problem. 
In PZSIC,  local orbitals are used to keep the corrections size-extensive. 
Traditionally, PZSIC requires solving the so called Pederson or localization equations (LE)\cite{doi:10.1063/1.446959,doi:10.1063/1.448266} to find the set of local orbitals that minimizes the total energy. 
Solving the LE and 
finding the optimal orbitals compliant with the condition is computationally expensive since it requires solving the LE for each pair of orbitals. 
Pederson \textit{et al.} in 2014  used 
Fermi-L\"owdin orbitals\cite{Luken1982,Luken1984} (FLOs) to solve the PZSIC equations. 
This approach is known as FLO-SIC\cite{doi:10.1063/1.4869581,PEDERSON2015153}. 
FLOs are L\"owdin orthogonalized set of Fermi orbitals (FOs) that can be obtained from the KS orbitals. 
The FOs depend on the density matrix and spin density.
 The FLOs are the local orbitals  that make PZSIC total energy  unitarily invariant.
For construction of FLOs, Fermi orbital descriptor (FOD) positions are used as $3N$ parameters in space 
that can be optimized in analogous manner to the  optimization of atomic positions 
in molecular structure optimization. FLOSIC method has 
computational advantage over traditional PZSIC since it requires optimizing only $3N$ parameters instead of $N^2$ parameters for the transformation to the local orbitals. 

Earlier applications of FLO-SIC with LSDA  showed significant improvements in atomic and molecular properties over SI-uncorrected LSDA performance\cite{doi:10.1021/acs.jctc.6b00112,doi:10.1063/1.4996498,kao2017role,FLOSICcode}.
Naturally, FLOSIC was later also applied to more sophisticated XC functionals than LSDA, such as 
Perdew–Burke-Ernzerhof (PBE) and Strongly Constrained and Appropriately Normed (SCAN),
 to see whether SIC improves the performance of those functionals in the higher rungs\cite{doi:10.1063/1.5125205,C9CP06106A,doi:10.1002/jcc.25767,doi:10.1063/1.5050809,Jackson_2019,PhysRevA.100.012505,doi:10.1021/acs.jpca.8b09940,doi:10.1063/1.5125205,doi:10.1063/1.5087065,doi:10.1063/1.5129533,doi:10.1021/acs.jctc.8b00344,Sharkas11283,doi:10.1063/1.5120532,SingHam,doi:10.1063/5.0004738,doi:10.1002/jcc.25586,schwalbe2019pyflosic}.
 PZSIC when applied to semi-local functional such as PBE GGA and SCAN meta-GGA provides good descriptions 
in stretched bond situation and provides bound atomic anions but this improvement occurs at the expense of 
worsening\cite{doi:10.1063/1.1794633,doi:10.1063/1.4752229,doi:10.1063/1.5087065,doi:10.1063/1.5120532,PhysRevA.84.050501,JONSSON20151858} 
the performance for properties where SI-uncorrected DFA performs well.
Shahi \textit{et al.}\cite{doi:10.1063/1.5087065} recently attributed the poor performance of 
PZSIC with GGAs and  higher rung functionals  to the nodality of the local orbital densities.
The use of complex localized orbitals with nodeless densities in PZSIC calculations by
Kl\"upfel, Kl\"upfel  and J\'onsson\cite{PhysRevA.84.050501} 
show that the complex orbital densities
alleviate the worsening of atomization energies when used with PBE functional.
This conflicting performance of PZSIC is called the paradox of SIC by Perdew and coworkers\cite{PERDEW20151}.
The worsening of energetics pertaining to equilibrium region primarily is a result of the overcorrecting
tendency of PZSIC. 
A few methods have been proposed to mitigate the overcorrecting tendency of PZSIC 
by scaling down the SIC contribution.
J\'onsson's group simply scaled the SIC by a constant scaling factor\cite{doi:10.1063/1.4752229}.
In a similar spirit, Vydrov \textit{et al.} proposed a method to scale down the SIC according to 
an orbital dependent scaling factor (OSIC)\cite{doi:10.1063/1.2176608}. This method however does 
not provide significant improvement over all properties. 
It improved over PZSIC atomization energies but worsened barrier heights. 
Moreover, the scaling approach by Vydrov \textit{et al.} results in 
worsening the asymptotic description of the effective potential causing atomic
anions to be unbound.  Ruzsinszky \textit{et al.}\cite{doi:10.1063/1.2387954} found that many-electron SIE and fractional-charge dissociation behavior of positively charged dimers reappear in the OSIC of Vydrov {\it et al.}.
A new selective OSIC method, called SOSIC, by Yamamoto and coworkers\cite{doi:10.1063/5.0004738} that selectively 
scales down the SIC in many electron regions overcomes the deficiencies of the OSIC method 
and gives stable atomic anions as well as improved total atomic energies. It also improves
the barrier heights over the OSIC method.
Very recently,  Zope {\it et al.}\cite{doi:10.1063/1.5129533} proposed a new SIC method
which identifies the single-electron
region using iso-orbital indicators and corrects for SIE in a pointwise fashion by scaling down the SIC. The iso-orbital indicator serves as a weight in numerical integration and identifies 
both the single-orbital regions where full correction is needed and the uniform density 
regions where the DFAs are already exact and correction is not needed. 
They called the new  SIC method local-SIC (LSIC)\cite{doi:10.1063/1.5129533} and assesed
its performance for a wide array of properties using LSDA. Unlike the PZSIC, the LSIC 
provided remarkable performance for both equilibrium properties like atomization 
energies and stretched bond situations that occur in barrier height calculation. 

The LSIC method makes use of iso-orbital indicator to identify one-electron region.
It offers additional degree of freedom in that suitable iso-orbital can be used 
or designed to identify one-electron region or tune the SIC contribution in a 
pointwise manner. In the original 
LSIC work, Zope {\it et al.} used a ratio of von Weisz\"acker and total
kinetic energy densities as a choice for the local scaling factor.
This iso-orbital indicator has been used in construction of self-correlation free 
meta-GGAs, in the regional SIC\cite{doi:10.1002/jcc.10279} and  also in 
local hybrid functionals\cite{jaramillo2003local,doi:10.1063/1.4865942}.
Several different choices for the local scaling factors are already available
in literature. Alternatively, new iso-orbital indicators particularly
suitable for LSIC can be constructed.
In this work, we explore the performance of the LSIC method using a simple 
ratio  of the orbital density and spin density as weight of 
SIC correction at a given point in space. This is the same scaling factor 
used by Slater to average the Hartree-Fock exchange potential in his 
classic work that introduced Hartree-Fock-Slater method\cite{slater1951simplification}.
We refer to this choice 
of scaling factor  as LSIC($w$) for the remainder of this manuscript and 
use LSIC($z$) to refer to the first LSIC application where the scaling factor is the ratio of von Weisz\"acker kinetic energy and kinetic energy densities. 
We investigate the performance of LSIC($w$) for  a few atomic properties: total energy, ionization potentials, and electron affinities. For molecules, we calculated the total energies, atomization energies, and the dissociation energies of a few selected systems.
We find that LSIC($w$) provides comparable results to LSIC($z$). 
We also show a case where LSIC($w$) performs better than the original LSIC($z$).
Additionaly, we examine the performance of the scaling factor $w$ based on the density ratio with the OSIC scheme. 

In the following section, brief descriptions of the  PZSIC, OSIC, and LSIC methods are presented. These methods
are implemented using the FLOs. Therefore, very brief definitions pertaining to FLOs are also presented.
The results and discussion are presented in the next sections.

\section{\label{sec:s2} Theory and computational method} 
\subsection{Perdew-Zunger and Fermi-Lowdin Self-Interaction Correction}
In the PZSIC method\cite{PhysRevB.23.5048}, SIE is removed on an orbital by orbital basis from the DFA energy as
\begin{equation}\label{eq:pzsic}
\begin{aligned}
  E^{PZSIC-DFA}&=E^{DFA}[\rho_{\uparrow},\rho_{\downarrow}]
  -\sum_{i\sigma}^{occ}\left\{ U[\rho_{i\sigma}]+E_{XC}^{DFA}[\rho_{i\sigma},0] \right\},
\end{aligned}
\end{equation}
where $i$ is the orbital index, $\sigma$ is the spin index, 
$\rho$ ($\rho_{i\sigma}$) is the electron density (local orbital density),
$U[\rho_{i\sigma}]$ is the exact self-Coulomb energy, and $E_{XC}^{DFA}[\rho_{i\sigma},0]$ is the self-exchange-correlation energy for a given DFA XC functional. 
Perdew and Zunger applied this scheme to atoms using the Kohn-Sham orbitals. For larger systems
the Kohn-Sham orbitals can be delocalized which would result in the violation of size extensivity.
Therefore local orbitals are required. This was recognized long ago by Slater and Wood\cite{slater1970statistical}
in 1971 and 
was also emphasized by Gopinathan\cite{PhysRevA.15.2135} in the context of 
self-interaction-correction of Hartree-Slater method
and later by Perdew and Zunger in the context of approximate Kohn-Sham calculations. 
Subsequent PZSIC calculations by Wisconsin group\cite{PhysRevB.28.5992,Harrison_1983,doi:10.1063/1.446959,doi:10.1063/1.448266,Harrison_1983}
used local  orbitals in variational implementation.
It was shown by Pederson and coworkers that local
orbitals used in the Eq. (\ref{eq:pzsic}) must satisfy the localization equations (LE) for variational  minimization of energy.
The LE for the orbitals $\phi_{i\sigma}$ is a pairwise condition and is given as
\begin{equation}\label{eq:LE}
\langle\phi_{i\sigma} |V_{i\sigma}^{SIC}-V_{j\sigma}^{SIC} | \phi_{j\sigma}\rangle=0.
\end{equation}

In the FLOSIC approach, FLOs are used in stead of directly solving the Eq. (\ref{eq:LE}).
First, FOs $\phi^{FO}$ are constructed with the density matrix and spin density at special positions in space called Fermi orbital descriptor (FOD) positions as
\begin{equation}\label{eq:3}
    \phi_{i }^{FO}(\vec{r}) = \frac{  \sum_{j}^{N_{occ}} \psi_{j}( \vec{a_{i}})\psi_{j}(\vec{r})   }   { \sqrt{\rho_{i}(\vec{a_{i}}) }}.
\end{equation}
Here, $i$ and $j$ are the orbital indexes, and $\psi$ is the KS orbital, $\rho_{i}$ is the electron spin density, and $\vec{a_{i}}$ is the FOD position. 
The FOs are then orthogonalized with the L\"owdin's scheme to form the FLOs. The FLOs are used for the calculation of the SIC energy and potential.
In this method, the optimal set of FLOs are found by finding the FODs that minimizes 
total energy. This optimization process is similar to that for geometry optimization.
We note that FLOs can be used in all three SIC (PZSIC, OSIC, and LSIC) methods.

\subsection{Orbitalwise scaling of SIC}
As mentioned in Sec. \ref{sec:introduction}, PZSIC tends to overcorrect the DFA energies and
several modifications to PZISC were proposed to \textit{scale down} the PZSIC correction.
 In the OSIC method of Vydrov \textit{et al}\cite{doi:10.1063/1.2176608} mentioned in Introduction
Eq. (\ref{eq:pzsic}) is modified to
\begin{equation}\label{eq:orbsic}
\begin{aligned}
    E^{OSIC-DFA}&=E_{XC}^{DFA}[\rho_\uparrow,\rho_\downarrow]
    -\sum_{i\sigma}^{occ}X_{i\sigma}^{k}\left(U[\rho_{i\sigma}]+E_{XC}^{DFA}[\rho_{i\sigma},0]  \right),
\end{aligned}
\end{equation}
where each local orbitalwise  
scaling factor $X_{i\sigma}^k$ is defined as
\begin{equation}\label{eq:OSIC_scalingfactor}
    X^{k}_{i\sigma}=\int 
    z_\sigma^k(\vec{r})
    \rho_{i\sigma}(\vec{r})d^3\vec{r}.
\end{equation}
Here, $i$ indicates the orbital index, $\sigma$ is the spin index, $z_\sigma$ is the iso-orbital indicator, and $k$ is an integer. 
The quantity $z_\sigma$ is used to interpolate the single-electron regions ($z_\sigma=1$) and uniform density region ($z_\sigma=0$).
In their original work, Vydrov \textit{et al.} used $z_\sigma = \tau_\sigma^W/\tau_\sigma$ to study the performance of OSIC with various XC functionals where $\tau_{\sigma}^W(\vec{r}) = |\vec{\nabla}\rho_{\sigma}(\vec{r})|^2/(8\rho_{\sigma}(\vec{r}))$ is the von Weisz\"acker kinetic energy density and $\tau_{\sigma}(\vec{r})=\frac{1}{2}\sum_i |\vec{\nabla}\psi_{i\sigma}(\vec{r})|^2$ is the non-interacting kinetic energy density.
Satisfying the gradient expansion in $\rho$ requires $k\geq1$ for LSDA, $k\geq2$ for GGAs, and $k\geq3$ for meta-GGA. Vydrov \textit{et al.}, however, used various values of $k$ to study its effect on the OSIC performance.

In their subsequent work, Vydrov \textit{et al.}\cite{doi:10.1063/1.2204599} used 
\begin{equation}\label{eq:rhoi_rho}
    w_{i\sigma}^k(\vec{r})=\left(\frac{\rho_{i\sigma}(\vec{r})}{\rho_\sigma(\vec{r})}\right)^k,
\end{equation}
the weight used by Slater in averaging Hartree-Fock potential, as a scaling factor instead of kinetic energy ratio.
They repeated the OSIC calculations using $w_{i\sigma}$ in place of $z_{\sigma}$ in Eq. (\ref{eq:OSIC_scalingfactor}).
Notice that Eq. (\ref{eq:rhoi_rho}) contains a local orbital index, this weight is thus an orbital 
dependent quantity.
$w_{i\sigma}$ approaches unity at single orbital regions since $\rho_\sigma(\vec{r}) = \rho_{i\sigma}(\vec{r})$ at this limit. Similarly, $w_{i\sigma}$ approaches zero at many-electron region since $\rho_\sigma(\vec{r}) \gg \rho_{i\sigma}(\vec{r})$ at this condition.
It was reported that the OSIC with Eq. (\ref{eq:rhoi_rho}) showed comparable performance as  
$z_\sigma = \tau_\sigma^W/\tau_\sigma$ despite of its simpler form.

\subsection{LSIC}
Though OSIC had some success in improving the performance with SIC, the approach leads 
to parameter $k$ dependent performance. Also, it gives  $-X_{HO}/r$ asymptotic potential instead of $-1/r$ 
for finite neutral systems and it results in inaccurate description of dissociation behavior\cite{doi:10.1063/1.2566637}.
In addition, many-electron SIE and fractional-charge dissociation behavior of positively charged dimers reemerge with the OSIC\cite{doi:10.1063/1.2387954}.
The recent LSIC method by Zope {\it et al.} applies the SIC in a different way than OSIC and retains
desirable beneficial features of PZSIC.
In LSIC, the SIC energy density is scaled down \textit{locally} as follows,
\begin{equation}\label{eq:LSIC}
\begin{aligned}
   E_{XC}^{LSIC-DFA}&= E_{XC}^{DFA}[\rho_{\uparrow},\rho_{\downarrow}] -\sum_{i\sigma}^{occ} \left( U^{LSIC}[\rho_{i\sigma}]   +E_{XC}^{LSIC}[\rho_{i\sigma},0]\right),
\end{aligned}
\end{equation}
where 
\begin{equation}\label{eq:LSIC_U}
     U^{LSIC}[\rho_{i\sigma}]=\frac{1}{2}\int d^3\vec{r}  
     \,z_\sigma(\vec{r})^k
     \,\rho_{i\sigma}(\vec{r})\int d^3\vec{r'}\,\frac{\rho_{i\sigma}(\vec{r'})}{|\vec{r}-\vec{r'}|},
\end{equation}
\begin{equation}\label{eq:LSIC_XC}
     E_{XC}^{LSIC}[\rho_{i\sigma},0]=\int d^3\vec{r} 
     \, z_\sigma(\vec{r})^k \,\rho_{i\sigma}(\vec{r}) \epsilon_{XC}^{DFA}([\rho_{i\sigma},0],\vec{r}).
\end{equation}
LSIC uses an iso-orbital indicator  to apply SIC pointwise in space.
An ideal choice of iso-orbital indicator should be such that LSIC reduces to DFA in uniform 
gas limit and reduces to PZSIC in the pure one-electron limit. To demonstrate the LSIC concept
Zope \textit{et al.} used  $z_\sigma=\tau_\sigma^W/\tau_\sigma$ as an iso-orbital indicator.
In this study, however, we use $w_{i\sigma}(\vec{r})
= \rho_{i\sigma}(\vec{r}) / \rho_{\sigma}(\vec{r})$ in place for $z_\sigma$ in Eqs. (\ref{eq:LSIC_U}) and (\ref{eq:LSIC_XC}).
We refer to the LSIC with $z_{\sigma}(\vec{r})$ as LSIC($z$) and LSIC with $w_{i\sigma}(\vec{r})$ as LSIC($w$) to differentiate the two cases.

\subsection{Computational details}
All of the calculations were performed using the developmental version of FLOSIC code\cite{FLOSICcode,FLOSICcodep}, a software based on the UTEP-NRLMOL code. 
PZSIC, OSIC, and LSIC methods using FLOs are implemented in this code.
FLOSIC/NRLMOL code uses Gaussian type orbitals\cite{PhysRevA.60.2840} whose default basis sets are 
in similar quality as quadruple zeta basis sets. 
We used the NRLMOL default basis sets throughout our calculations.  
For calculations of atomic anions, long range s, p, and d single Gaussian orbitals are 
added to give a better description of the extended nature of anions. 
The exponents $\beta$ of these added single Gaussians were obtained using the relation, $\beta(N+1)=\beta(N)^2/\beta(N-1)$,
where $N$ is the $N$-th exponent. FLOSIC code uses a variational integration mesh\cite{PhysRevB.41.7453} that provides accurate numerical integration.

In this work, our focus is on the LSDA functional because LSIC applied to LSDA is free from the gauge problem\cite{doi:10.1063/5.0010375} unlike GGAs and meta-GGAs where a gauge transformation is needed since their XC potentials are not in the Hartree gauge.
We used an SCF energy convergence criteria of $10^{-6}$
Ha for the total energy and an FOD force tolerance of $10^{-3}$ Ha/bohr for FOD optimizations in FLOSIC calculations.
For OSIC and LSIC calculations, we used respective FLOSIC densities and FODs as a starting point and performed a non-self-consistent  
calculation of energy on the FLOSIC densities. Several values for the scaling power $k$ are
used in the LSIC($w$) and OSIC($w$) calculations.
The additional computational cost of the scaling factor in OSIC and LSIC is very small 
compared to a regular FLO-PZSIC calculation.

\section{\label{sec:results} Results and discussion}
The LSIC method was assessed for a wide array of electronic structure properties to 
obtain a good understanding of how the new methodology performs. Here, 
we asses the performance of LSIC($w$) vis-a-vis  LSIC($z$) and OSIC($w$) 
using the same array of electronic properties. 
We considered total energies, ionization potentials, and electron affinities for atoms and atomization energies, reaction barrier heights, and dissociation energies for molecules.

\subsection{\label{sec:s3s1} Atoms }
In this section, we present our results on   
total energies, ionization potentials, 
and electron affinities for atoms.

\subsubsection{\label{sec:s3s2} Total energy of atoms}
\begin{figure}
    \centering
    \includegraphics[width=0.8\columnwidth]{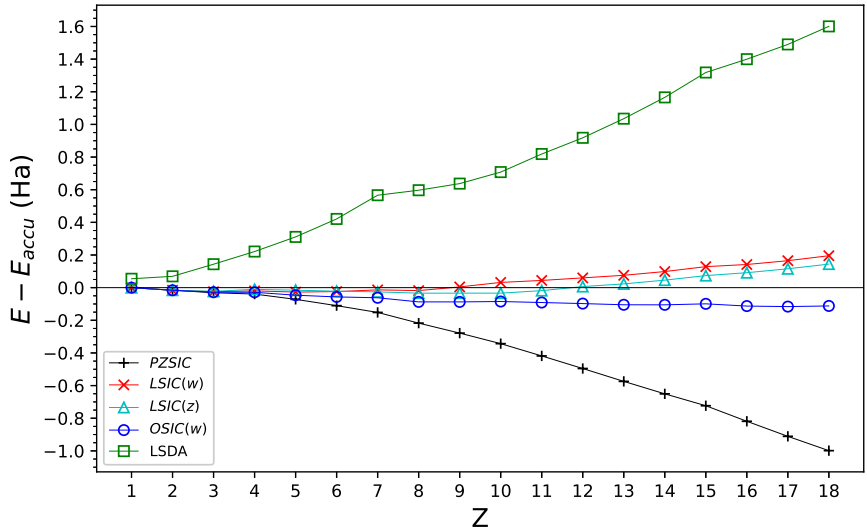}
    \caption{Total energy difference (in hartree) of atoms $Z=1-18$ with respect to accurate nonrelativistic estimates\cite{PhysRevA.47.3649}.}
    \label{fig:atoms-diff}
\end{figure}

We compared the total atomic energies of the atoms $Z=1-18$ against accurate non-relativistic values reported by Chakravorty \textit{et al.}\cite{PhysRevA.47.3649}. Various integer values of $k$ were used for LSIC($w$) and OSIC($w$).
The differences between our calculated total energies with $k=1$ and the reference 
values are plotted in Fig. \ref{fig:atoms-diff}.  The plot clearly shows the effect 
of scaling on the total energies of atoms. Consistent with reported results, the
LSDA total energies are too high compared to accurate reference values\cite{PhysRevA.47.3649} whereas PZSIC consistently underestimates the total energies due 
to its over correcting tendency. The LSIC method, where both scaling factors
performs similarly, provides the total energies closer to the reference values than LSDA and PZSIC-LSDA. 
Likewise, OSIC method also reduces the overcorrection bringing
the total energies to close agreement with the reference values. 
The  mean absolute errors  (MAEs) in total energy with respect to the reference 
for various $k$ values are shown in Table \ref{tab:table1}. The MAE of PZSIC is $0.381$ Ha whereas
LSIC($w$) and OSIC($w$) show MAEs of $0.061$ and $0.074$ Ha, respectively, with $k=1$. 
LSIC($z$) shows a better performance than OSIC($w$) and LSIC($w$). 
The LSIC($w$) MAE is in the same order of magnitude as the earlier reported MAE of
LSIC($z$) of 0.041 Ha\cite{doi:10.1063/1.5129533}.
As the value of $k$ increases, the magnitude of SI-correction is reduced. This result in MAEs become larger for $k > 1$ eventually approaching the LSDA numbers.

For $k=0$ the scaled methods correctly produce the PZSIC results. The scaling is 
optimal for $k=1$ which results in 
optimal magnitude of SI-correction for LSIC($w$) and almost right magnitude for OSIC($w$).
The magnitude of SIC energy of each orbitals when compared among 
different methods, it is found that the SIC correction in  LSIC($w$) is larger (i.e. less scaling down) for
the core orbitals than in the LSIC($z$). This trend is reversed for the valence orbitals 
(cf. Table \ref{table:sic_amount}).  It can be seem from Table \ref{table:sic_amount} that 
total SIC energy in both methods is essentially similar in magnitude. However the 
way scaling factors behave affects the orbitalwise contribution to the total SIC 
energy. This changes the SIC potentials and results in  two methods 
performing differently for cations and anions.
For OSIC($w$), we find the smallest MAE for $k=2$ of $0.070$ Ha, a value slightly smaller than that for $k=1$.

\begin{table}
\caption{\label{tab:table1}Mean absolute error of the total atomic energy (in hartree) for atoms $Z=1-18$ 
with respect to accurate nonrelativistic estimates\cite{PhysRevA.47.3649}. 
}
\begin{tabular*}{0.68\textwidth}{@{\extracolsep{\fill}}lc}
\toprule
Method &MAE\\
\midrule
PZSIC           & 0.381\\
LSIC($z, k=1$)  & 0.041\\\hline
LSIC($w, k=1$) & 0.061\\
LSIC($w, k=2$) & 0.196\\
LSIC($w, k=3$) & 0.277\\
LSIC($w, k=4$) & 0.332\\ \hline    
OSIC($w, k=1$) & 0.074\\
OSIC($w, k=2$) & 0.070\\
OSIC($w, k=3$) & 0.135\\
\bottomrule
\end{tabular*}
\end{table}

\begin{table}
\caption{\label{table:sic_amount}Magnitude of SIC energy (in hartree) per orbital type in Ar atom for each method.}
\begin{tabular*}{0.68\textwidth}{@{\extracolsep{\fill}}ccccc}
\toprule
Orbital & PZSIC & LSIC($z$) & LSIC($w$) & OSIC($w$)\\
\midrule
1s	    &-0.741	&-0.387	&-0.490	&-0.584 \\
2sp$^3$	&-0.126	&-0.070	&-0.050	&-0.062 \\
3sp$^3$	&-0.016	&-0.017	&-0.006	&-0.008 \\
\midrule
Total SIC	&-2.616	&-1.473	&-1.421	&-1.729 \\
\bottomrule
\end{tabular*}
\end{table}

\subsubsection{\label{sec:s3s3} Ionization potential}
\begin{figure}
    \centering
    \includegraphics[width=0.8\columnwidth]{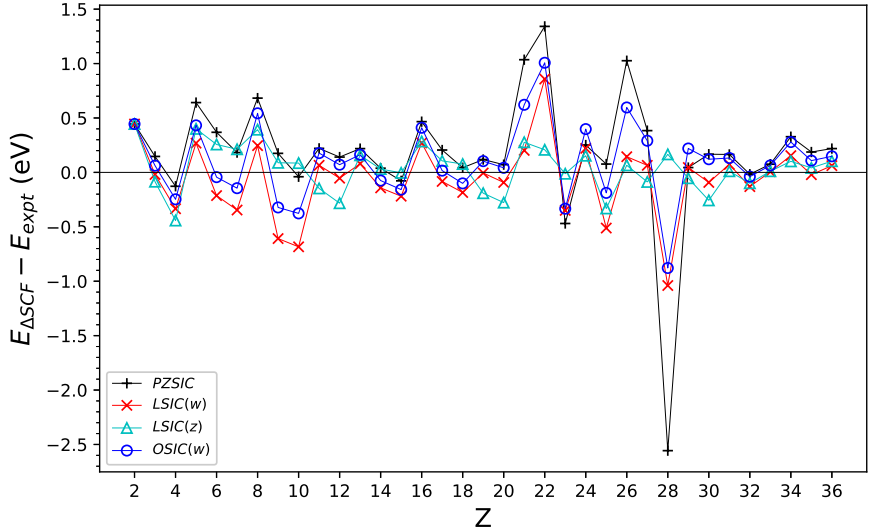}
    \caption{Energy difference in ionization potential (in eV) for a set of atoms $Z=2-36$ with respect to experimental values\cite{NIST_ASD}.}
    \label{fig:atoms-IP}
\end{figure}

The ionization potential (IP) is the energy required to remove an electron from the outermost orbital. 
Since electron removal energy is related to the asymptotic shape of the potential, one can expect SIC plays an important role in determining IPs. 
We calculated the IPs using the $\Delta$SCF method defined as
\begin{equation}
    E_{IP}=E_{cat}-E_{neut}
\end{equation}
where $E_{cat}$ is the total energy in the cationic state and $E_{neut}$ is the total energy at the neutral state. 
The calculations were performed for atoms from helium to krypton, and we compared the computed IPs against the experimental ionization energies \cite{NIST_ASD}.
FODs were relaxed both for neutral atoms and for their cations.
Fig. \ref{fig:atoms-IP} shows the difference of calculated IPs with respect to the reference values.
MAEs with different methods are shown in 
Table \ref{tab:table2}
for a subset $Z=2-18$ as well as for the entire set $Z=2-36$ to facilitate a comparison against literature.
For the smaller subset, $Z=2-18$,
the MAEs are $0.248$ and $0.206$ eV for PZSIC and LSIC($z$), respectively. 
The MAE for OSIC($w$, $k=1$) is $0.223$ eV showing an improvement over PZSIC.
LSIC($w$, $k=1$) shows MAE of $0.251$ eV, a comparable error with PZSIC.
MAEs increase for LSIC($w$, $k\geq2$) and OSIC($w$, $k\geq2$) in comparison to their respective $k=1$ MAEs. 
Interestingly, however, when we considered the entire set of atoms ($Z=2-36$), LSIC($w$) has
MAEs of $0.238$ and $0.216$ eV for $k=1$ and $k=2$ respectively showing smaller errors than 
PZSIC (MAE, $0.364$ eV) but LSIC($w$) falls short of LSIC($z$) which has the smallest error (MAE, 0.170 eV). 
For this case, OSIC($w$, $k=1-3$) shows better performance than PZSIC but not as well as LSIC($w$)
for a given $k$.
LSIC($z$) performs better than both LSIC($w$) and OSIC($w$).
The difference in performance between LSIC($z$) and LSIC($w$) implies 
that scaling of SIC for the cationic states is more sensitive to the choice of local scaling factor
than for the neutral atoms.

\begin{table}
\caption{\label{tab:table2} Mean absolute error of ionization potentials (in eV) for set of atoms $Z=2-18$ and $Z=2-36$ with respect to experiment\cite{NIST_ASD}.}
\begin{tabular*}{0.68\textwidth}{@{\extracolsep{\fill}}lcc}
\toprule
Method &Z=2-18 (17-IPs) & Z=2-36 (35-IPs)\\
\midrule
PZSIC              & 0.248 & 0.364\\
LSIC($z, k=1$)  & 0.206 & 0.170 \\\midrule
LSIC($w, k=1$) & 0.251 & 0.238 \\
LSIC($w, k=2$) & 0.271 & 0.216\\
LSIC($w, k=3$) & 0.297 & 0.247\\
LSIC($w, k=4$) & 0.324 & 0.284\\\hline
OSIC($w, k=1$) & 0.223 & 0.267\\
OSIC($w, k=2$) & 0.247 & 0.247\\
OSIC($w, k=3$) & 0.255 & 0.259\\
\bottomrule
\end{tabular*}
\end{table}

\subsubsection{\label{sec:s3s4} Electron affinity }
The electron affinity (EA) is the energy released when an electron is added to the 
system. We studied EAs for 20 atoms that are experimentally found to bind an 
electron\cite{NIST_CCCBD}. They are H, Li, B, C, O, F, Na, Al, Si, P, S, Cl, K, Ti, 
Cu, Ga, Ge, As, Se, and Br atoms. The EAs were calculated using the $\Delta$SCF 
method $E_{EA}=E_{neut}-E_{anion}$ and values were compared against the 
experimental EAs\cite{NIST_CCCBD}.

\begin{figure}
    \centering
    \includegraphics[width=0.8\columnwidth]{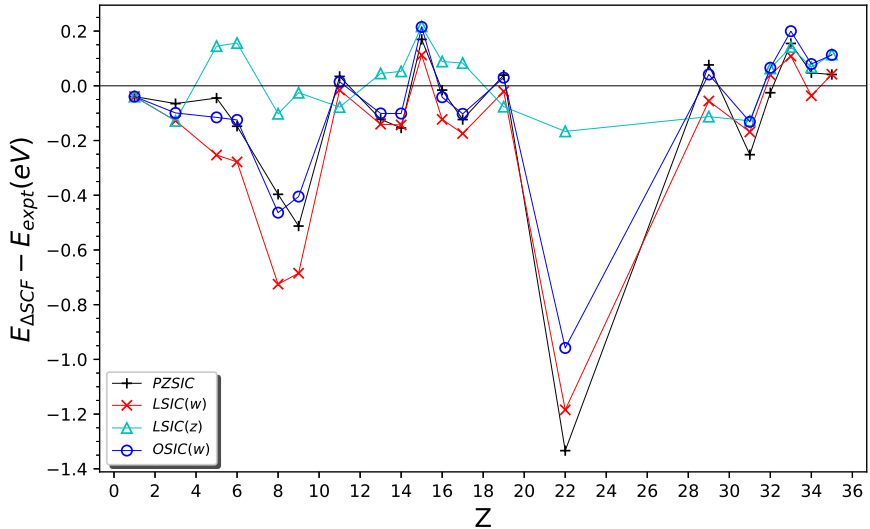}
    \caption{Electron affinity (eV) difference for atoms $Z=2-36$ with respect to experiment\cite{NIST_CCCBD}.}
    \label{fig:atoms-EA}
\end{figure}

Fig. \ref{fig:atoms-EA} shows deviation of EA from reference experimental 
values for various methods. The MAEs are summarized in Table \ref{tab:table3}.
We have presented the MAEs in two sets, the smaller subset which
contains hydrogen through chlorine (12 EAs) and for the complete set, hydrogen to bromine (20 EAs).

For 12 EAs, MAEs for PZSIC and LSIC($z$) are $0.152$ and $0.097$ eV, respectively. OSIC($w$) shows MAE of $0.152$ eV for $k=1$, the same performance as PZSIC. LSIC($w$), however, does not perform as well as PZSIC, giving the MAEs of $0.235$ eV for $k=1$. In both case, the error decrease slightly for $k\geq2$ but there is no significant impact on their performance.

For 20 EAs, the similar trend persists. PZSIC and LSIC($z$) have MAEs of $0.190$ and $0.102$ eV, respectively. The MAEs of LSIC($w$) are in the range $0.176-0.224$ eV for $k=1-4$ and those of OSIC($w$) are between $0.155-0.172$ eV for $k=1-3$. Again, decrease in error is observed as the value in $k$  increases.  
In particular,  larger discrepancy between LSIC($w$,$k=1$) and experiment is seen for O, F, and Ti
atoms. This is due to LSIC($w$) raising the anion energies more than their neutral state energies.

\begin{table}
\caption{\label{tab:table3}Mean absolute error in electron affinities (in eV) for 12 EAs and 20 EAs set of atoms with respect to experiment\cite{NIST_CCCBD}.}
\begin{tabular*}{0.68\textwidth}{@{\extracolsep{\fill}}lcc}
\toprule
Method &(12 EAs) MAE & (20 EAs) MAE \\
\midrule
PZSIC              & 0.152 & 0.190\\
LSIC($z, k=1$)  & 0.097 & 0.102 \\\midrule
LSIC($w, k=1$) & 0.235 & 0.224\\
LSIC($w, k=2$) & 0.229 & 0.205\\
LSIC($w, k=3$) & 0.215 & 0.189\\
LSIC($w, k=4$) & 0.202 & 0.176\\
\midrule
OSIC($w, k=1$) & 0.152 & 0.172\\
OSIC($w, k=2$) & 0.150 & 0.164\\
OSIC($w, k=3$) & 0.145 & 0.155\\
\bottomrule
\end{tabular*}
\end{table}

\subsection{\label{sec:s3s5} Atomization energy}
\begin{figure}
    \centering
    \includegraphics[width=0.68\columnwidth]{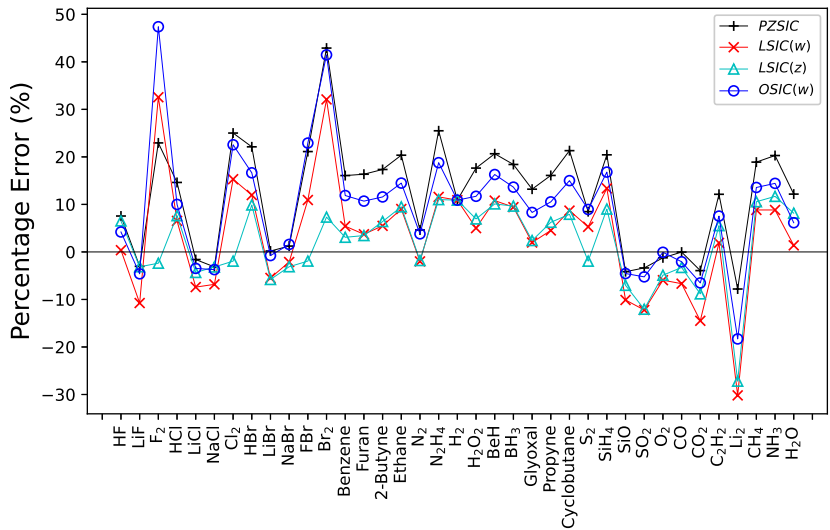}
    \caption{Percentage errors of atomization energy (\%) for a set of 37 molecules with respect to experimental values\cite{NIST_CCCBD} using different scaling methods.}
    \label{fig:atoms-AE}
\end{figure}

To study the performance of LSIC($w$) for molecules, first, we calculated the atomization energies (AEs) of 37 selected molecules. Many of these molecules are subset of the G2/97 test set\cite{doi:10.1063/1.460205}. The 37 molecules set includes systems from the AE6 set\cite{doi:10.1021/jp035287b}, small but a good representative of the main group atomization energy (MGAE109) set\cite{doi:10.1063/1.3663871}.
The AEs were calculated by taking the energy difference of fragment atoms 
and the complex, that is, $ AE=\sum_{i}^{N_{atom}}E_i-E_{mol}>0.$
$E_i$ is the total energy of an atom, $E_{mol}$ is the total energy of the molecule, and $N_{atom}$ is the number of atoms in the molecule. The calculated AEs were compared to the non-spin-orbit coupling reference values\cite{doi:10.1063/1.3663871} for AE6 set and to the experimental values\cite{NIST_CCCBD} for the entire set of 37 molecules.
The percentage errors obtained through various methods are shown in Fig. \ref{fig:atoms-AE}.  
The overestimation of AEs with PZSIC-LSDA due to overcorrection is  rectified in LSIC($w$).
We have summarized  MAEs and mean absolute percentage errors (MAPEs) of AE6 
and 37 molecules from G2 set in
Table \ref{tab:table4}.
For AE6 set, MAEs for PZSIC, LSIC($z$), LSIC($w,k=1$), and 
OSIC($w,k=1$) are $57.9$, $9.9$, $13.8$, and $33.7$ kcal/mol respectively.
The MAE in  LSIC($z$) is about 4 kcal/mol larger than LSIC($w,k=1$) 
but substantially better than the PZSICs or OSIC($w$). 
For the larger $k$ in LSIC($w$), however, the performance starts to degrade 
with consistent increase in the MAE of $33.5$ kcal/mol for $k=4$. This is in 
contrast to  OSIC where the performance improves for $k=2$ and $3$ compared to
$k=1$. The scaling thus affect differently in the two methods. 
OSIC($w,k=1$) tends to slightly underestimate total energies.
By increasing $k$, total energies shift toward the LSDA total energies 
and improves performance for moderate increase in $k$. 
On the contrary, total energies are slightly overestimated for LSIC($w,k=1$), 
and increasing $k$ makes the energies deviate away from the accurate estimates. 
OSIC($w,k=3$) and LSIC($w,k=1$) have a similar core orbital SIC energy.
In their study of OSIC($w$), Vydrov and Scuseria\cite{doi:10.1063/1.2204599} used  values 
of $k$ up to 5 and found the smallest error of $k=5$ (MAE, $11.5$ kcal/mol).  
But we expect the OSIC performance to degrade eventually 
for large $k$ since increase in $k$ results in increase in 
quenching of the SIC correction  thus the results will eventually 
approach to those of DFA, in this case LSDA.
For the full set of 37 molecules, PZSIC, LSIC($z$),  LSIC($w,k=1$), and OSIC($w,k=1$) 
show the MAPEs of $13.4$, $6.9$, $9.5$ and $11.9$\%, respectively. 
OSIC($w$) shows a slight improvement in MAPE for $k=2$ and $3$. 
For the larger set, LSIC($w$) consistently shows smaller MAPEs than OSIC($w$) for $k=1-3$.
All four values of $k$ with the LSIC($w$) in this study showed better performance than PZSIC for the 37 molecules set.

\begin{table}
\caption{\label{tab:table4}Mean absolute error (in kcal/mol) and mean absolute percentage error (in \%) of atomization energy for AE6 set of molecules
\cite{doi:10.1063/1.3663871} 
and for the set of 37 molecules from G2 set with respect to experiment\cite{NIST_CCCBD}.}
\begin{tabular*}{0.68\textwidth}{@{\extracolsep{\fill}}lccc}
\toprule
       & AE6 MAE     &    AE6   & 37 molecules \\
Method & (kcal/mol)   &  MAPE (\%) & MAPE (\%) \\
\midrule
PZSIC          & 57.9 & 9.4 & 13.4\\
LSIC($z, k=1)$ &  9.9 & 3.2 &  6.9\\\midrule
LSIC($w, k=1$) & 13.8 & 4.4 &  9.5\\
LSIC($w, k=2$) & 18.6 & 5.3 &  9.1\\
LSIC($w, k=3$) & 26.9 & 5.8 &  9.2\\
LSIC($w, k=4$) & 33.5 & 6.7 &  9.7\\\midrule
OSIC($w, k=1$) & 33.7 & 6.3 &  11.9\\
OSIC($w, k=2$) & 24.1 & 5.1 &  11.3\\
OSIC($w, k=3$) & 17.8 & 4.3 &  10.9\\
\bottomrule
\end{tabular*}
\end{table}

\subsection{\label{sec:s3s6} Barrier heights }
Accurate description of  chemical reaction barrier is challenging for DFAs since it involves calculation of 
energies in  non-equilibrium situations. 
In most of the cases, the saddle point energies are underestimated since DFAs do 
not perform well for a non-equilibrium state that involves a stretched bond.
This shortcoming of DFAs in a stretched bond case arises from SIE; when an electron is shared
and stretched out, SIE incorrectly lowers the energy of transition state. 
SIC handles the stretched bond states accurately and provides a correct picture in chemical reaction paths. 
We studied the reaction barriers using the BH6\cite{doi:10.1021/jp035287b} set of molecules for LSIC($w$) method. BH6 is a representative subset of the larger BH24\cite{doi:10.1021/ct600281g} set consisting of three reactions OH + CH$_4$ $\rightarrow$ CH$_3$ + H$_2$O,  H + OH $\rightarrow$ H$_2$ + O, and H + H$_2$S  $\rightarrow$ H$_2$ + HS. 
We calculated the total energies of left- and right-hand side and at the saddle point of these chemical reactions. The barrier heights for the forward (f) and reverse (r) reactions were obtained by taking the energy differences of their corresponding reaction states. 

The mean errors (MEs) and MAEs of  computed barrier heights against the reference values\cite{doi:10.1021/jp035287b} 
are compared in Table \ref{tab:table5}.
MAEs for PZSIC, LSIC($z$), LSIC($w, k=1$), and OSIC($w,k=1$) are $4.8$, $1.3$, $3.6$, and $3.6$ kcal/mol, respectively.
PZSIC significantly improves MAE compared to LSDA (MAE, 17.6 kcal/mol),
LSIC($w, k=1$) further reduces the error from PZSIC. Its ME and MAE indicate that there is no 
systematic underestimation or overestimation. LSIC($w, k=1$) also further improves the
PZSIC numbers but not to the same level as LSIC($z$).
For $k\geq2$, MAEs increases systematically for LSIC($w, k\geq2$) 
though small MEs are seen for LSIC($w, k=2,3$).  The performance deteriorates for $k>2$ 
beyond that of PZSIC. 
OSIC($w$) shows marginally better performance  than PZSIC.
Vydrov and Scuseria\cite{doi:10.1063/1.2204599} showed that the best performance is achieved with $k=1$ (MAE, $3.5$ kcal/mol). 
The performance improvement with OSIC is not as dramatic as LSICs in terms of MEs and MAEs 
where the rather large MEs are seen i
Overall LSIC($w$) performs better than OSIC($w$) for barrier heights.

\begin{table}
\caption{\label{tab:table5}Mean error (in kcal/mol) and mean absolute error (in kcal/mol) of BH6 sets of chemical reactions\cite{doi:10.1021/jp035287b}.}
\begin{tabular*}{0.68\textwidth}{@{\extracolsep{\fill}}lccc}
\toprule
Method & ME (kcal/mol) & MAE (kcal/mol) \\
\midrule
%LSDA & -17.6 & 17.6 \\
PZSIC              & -4.8  &  4.8 \\
LSIC($z, k=1$)  &  0.7  &  1.3 \\\midrule
LSIC($w, k=1$) & -1.0  &  3.6 \\
LSIC($w, k=2$) & -0.1  &  4.6 \\
LSIC($w, k=3$) &  0.3  &  5.0 \\
LSIC($w, k=4$) &  0.6  &  5.5 \\\midrule
OSIC($w, k=1$) & -3.4  &  3.6   \\
OSIC($w, k=2$) & -3.1  &  4.1 \\
OSIC($w, k=3$) & -3.0  &  4.6 \\
\bottomrule
\end{tabular*}
\end{table}

\subsection{\label{sec:s3s7}Dissociation and reaction energies}
A pronounced manifestation of SIE is seen in dissociation of positively charged dimers $X_2^+$. 
SIE causes the system to dissociate into two fractionally charges cations  
instead of $X$ and $X^{+}$. Here we use 
the SIE4x4\cite{C7CP04913G} and SIE11\cite{doi:10.1021/ct900489g} sets to study 
the performance of LSIC($w$) and OSIC($w$) in correcting the SIEs.
The SIE4x4 set consists of dissociation energy calculations of four positively 
charged dimers at varying bond distances $R$ from their equilibrium distance 
$R_e$ such that $R/R_e$ = 1.0, 1.25, 1.5 and 1.75.
The dissociation energy $E_D$ is calculated as
\begin{equation}
    E_D=E(X)+E(X^+)-E(X_2^+).
\end{equation}
The SIE11 set consists of eleven reaction energy calculations: five cationic reactions and six neutral reactions.
These two sets are commonly used for studying the SIE related problems.
The calculated dissociation and reaction energies are compared against the CCSD(T) reference values\cite{C7CP04913G,doi:10.1021/ct900489g}, and MAEs 
are summarized in Table \ref{tab:table6}.
For the SIE4x4 set, PZSIC, LSIC($z$), LSIC($w, k=1$), and OSIC($w,k=1$) show MAEs of $3.0$, $2.6$, $4.7$ and $5.2$ kcal/mol. 
LSIC($z$) provides small improvement in equilibrium energies while keeping accurate behavior of PZSIC at the dissociation limit resulting in marginally better performance. 
LSIC($w$) shows errors a few kcal/mol larger than PZSIC. 
This increase in error arises 
because LSIC($w$) alters the (NH$_3$)$_2^+$ and (H$_2$O)$^+_2$ dissociation curves. 
In LSIC($z$) the scaling of SIC occurs mostly for the core orbitals  (Cf. Table \ref{table:sic_amount})
whereas LSIC($w$) also includes some noticeable scaling down effect from valence orbitals.
This different scaling behavior seems to contribute to different dissociation curves.
Lastly, OSIC($w$) has a slightly larger error than LSIC($w$).

For the SIE11 set, MAEs are $11.5$, $4.5$, $8.3$, and $11.1$ kcal/mol for PZSIC, LSIC($z$), LSIC($w, k=1$), and OSIC($w,k=1$), respectively.
All scaled-down approaches we considered, LSIC($z$) and LSIC($w$), and OSIC($w$) showed performance 
improvement over PZSIC.
LSIC($z$) shows the largest error reduction  
by 60\%, while  LSIC($w, k=1$) shows 28\%  decrease in error with respect to PZSIC. 
OSIC($w$) with $k=1-3$ has slightly smaller MAEs within 1 kcal/mol of PZSIC. 
LSIC($z$) method improves cationic reactions more than neutral reactions with respect to PZSIC.
Increase in $k$ beyond 2 results in too much suppression of SIC and leads to increase in error for LSIC($w, k\geq2$).
LSIC($w$) yielded consistently smaller MAEs than OSIC($w$) but larger than LSIC($z$) over the whole SIE11 reactions.

\begin{figure}
   \centering
   \includegraphics[width=0.8\columnwidth]{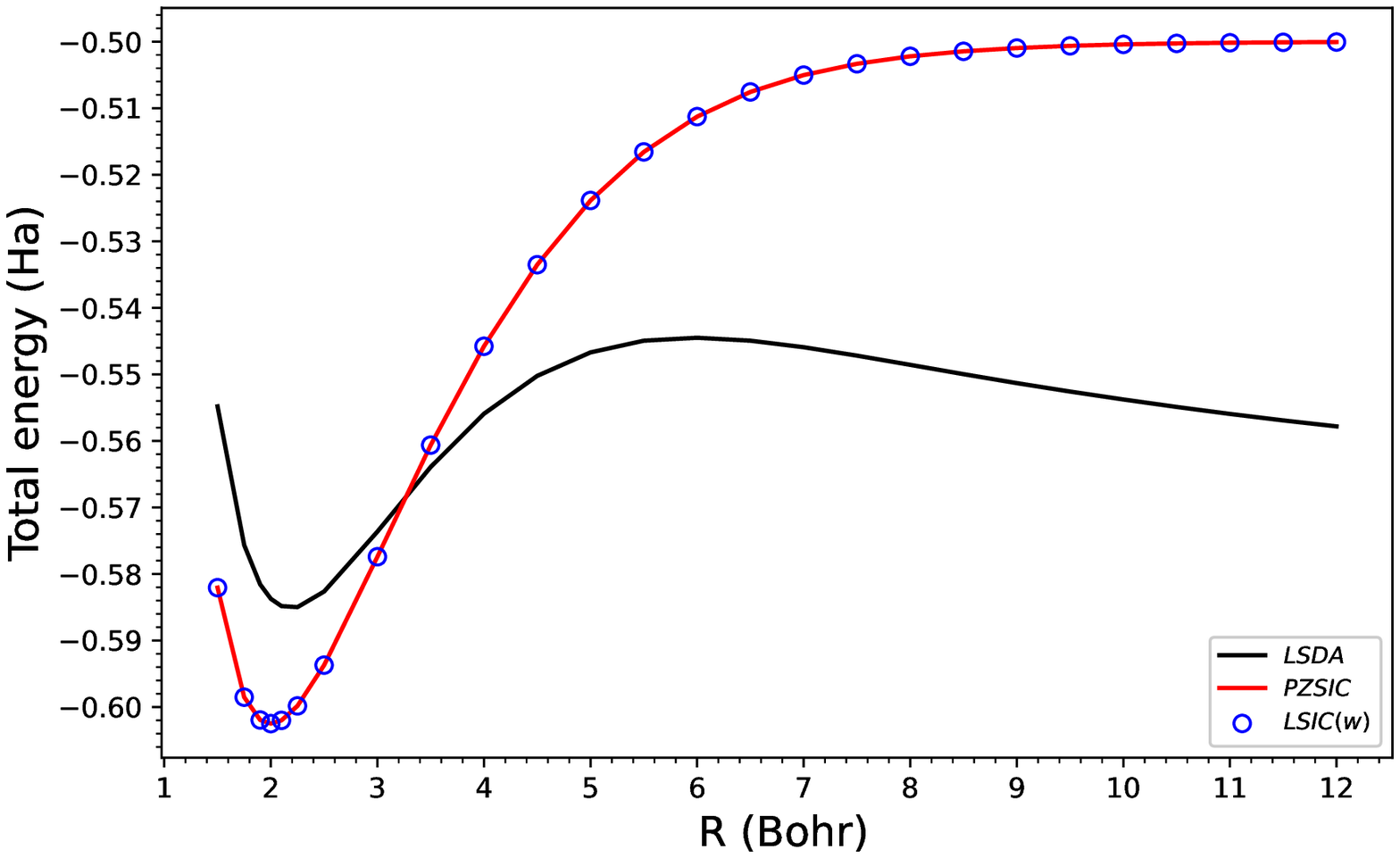}
   
   \includegraphics[width=0.8\columnwidth]{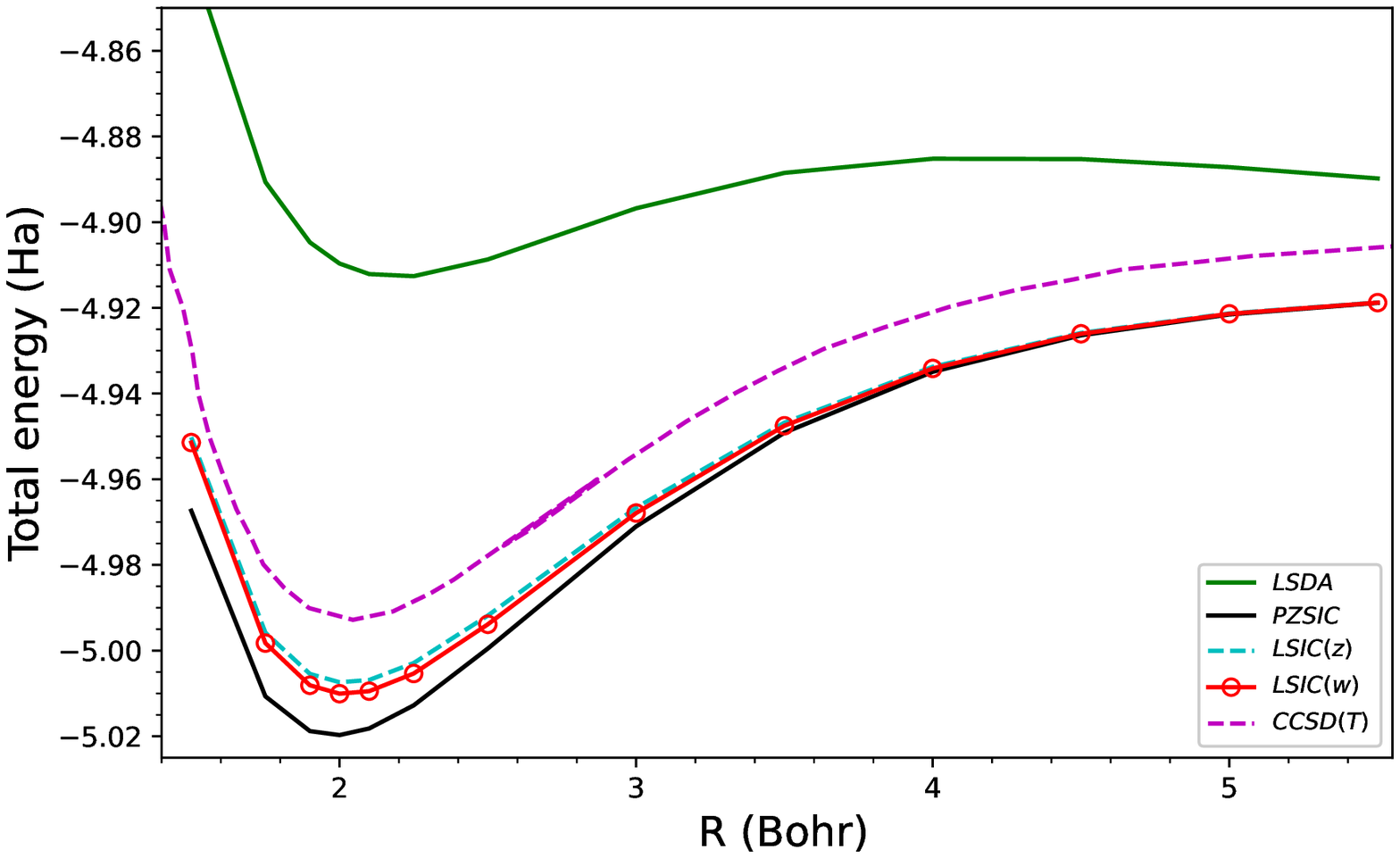}
   \caption{Dissociation curves of (a)  H$_2^+$ and (b)  He$_2^+$ using various methods. The CCSD(T) curve from Ref.~[\onlinecite{doi:10.1063/1.2566637}] is plotted for comparison.}
   \label{fig:dissociation}
\end{figure}

Finally,
we show the ground-state dissociation curves for H$^+_2$ and He$^+_2$ in Fig. \ref{fig:dissociation}. As previously discussed in literature \cite{doi:10.1021/jp0534479}, 
DFAs at large separation cause the complexes 
to dissociate into two 
fragments atoms.
PZSIC restores the correct dissociation behavior at the large separation distance.
When LSIC is applied, the behavior of PZSIC at the dissociation limit is preserved in both LSIC($z$) and present 
LSIC($w$).
For H$_2^+$, a one-electron system, LSIC reproduces the identical behavior as PZSIC [Fig. \ref{fig:dissociation} (a)].
For He$_2^+$, a three-electron system, LSIC applies the correction to PZSIC only near equilibrium regime
[Fig. \ref{fig:dissociation} (b)]. 
LSIC brings the equilibrium energy closer to the CCSD energy compared to PZSIC energy.
The implication of Fig. \ref{fig:dissociation} is that the present scaling factor $w$ performs well in differentiating the single-orbital like regions and many-electron like regions.

\begin{table}
\caption{\label{tab:table6} Mean absolute error for dissociation and reaction energies (in kcal/mol) of SIE4x4 and SIE11 sets of chemical reactions with respect to CCSD(T)\cite{C7CP04913G,doi:10.1021/ct900489g}.}
\begin{tabular*}{0.68\textwidth}{@{\extracolsep{\fill}}lcccc}
\toprule
Reaction & SIE4x4  &  SIE11 & SIE11 & SIE11 \\
    &  &   & 5 cationic & 6 neutral \\
\hline
PZSIC              & 3.0 & 11.5 & 14.9 & 8.7 \\
LSIC($z$)       & 2.6 &  4.5 &  2.3 & 6.3 \\\midrule
LSIC($w$) (k=1) & 4.7 &  8.3 &  8.6 & 8.0 \\
LSIC($w$) (k=2) & 5.5 &  8.3 &  8.3 & 8.3 \\
LSIC($w$) (k=3) & 5.8 &  8.8 &  8.2 & 9.3 \\
LSIC($w$) (k=4) & 5.9 &  9.3 & 8.2 & 10.2\\\midrule
OSIC($w$) (k=1) & 5.2 & 11.1 & 13.7 & 9.0 \\
OSIC($w$) (k=2) & 6.0 & 11.0 & 13.5 & 9.0 \\
OSIC($w$) (k=3) & 6.4 & 10.9 & 13.3 & 8.8\\
\bottomrule
\end{tabular*}
\end{table}

\subsection{ Water binding energies: a case where LSIC($z$) performs poorly}
\begin{figure}
    \centering
    \includegraphics[width=0.68\columnwidth]{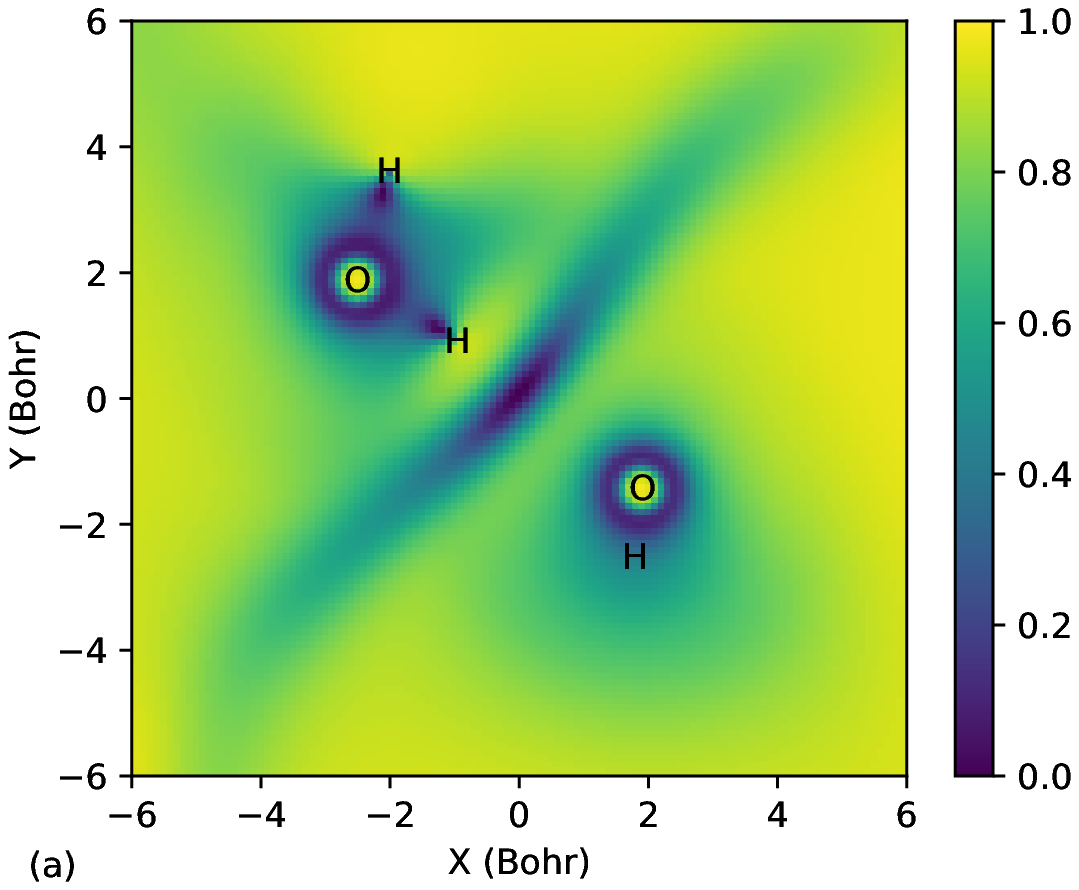}
    
    \includegraphics[width=0.68\columnwidth]{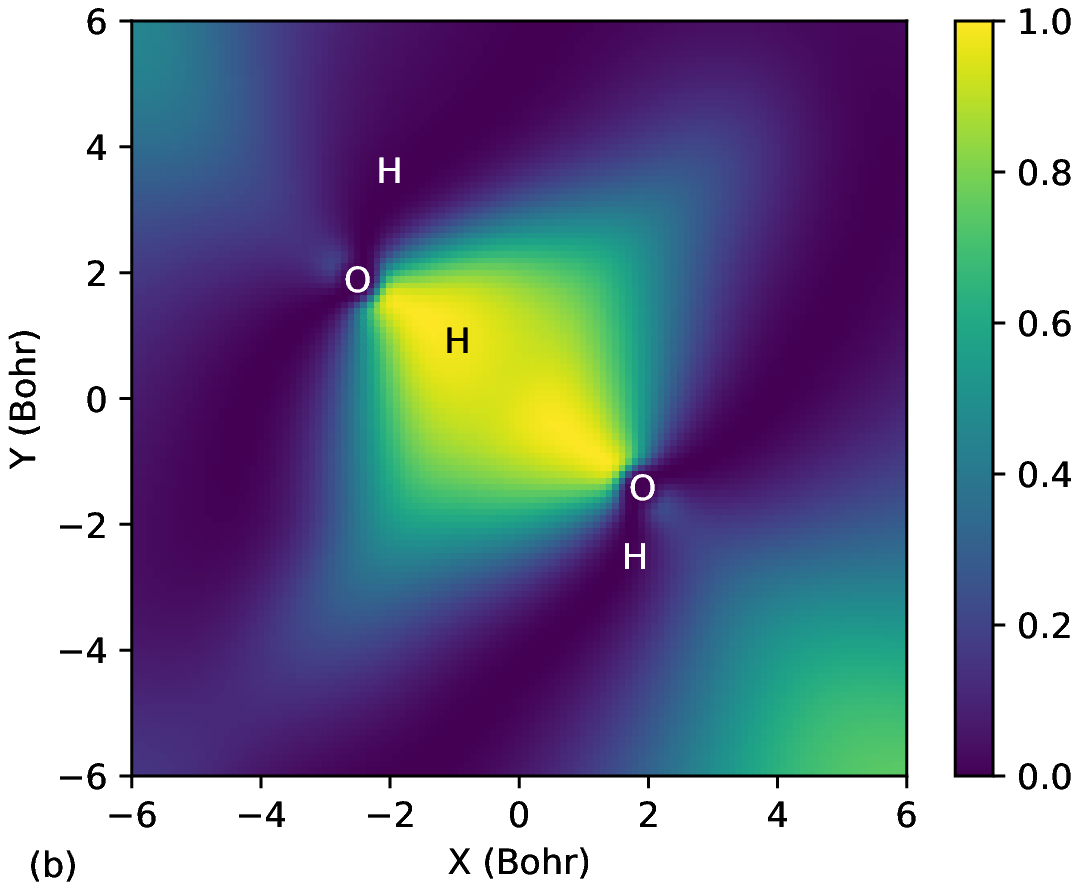}
    \caption{Cross sectional plots of the iso-orbital indicators for water cluster dimer: (a) $\tau^W/\tau$ and (b) $\rho_i/\rho$'s from the two FLOs along the hydrogen bond.}
    \label{fig:waterbind}
\end{figure}

Kamal \textit{et al.}\cite{Sharkas11283} recently studied binding energies of small water 
clusters using the PZSIC method in conjunction with FLOs to examine the effect of SIC on 
binding energies of these systems. Water clusters are bonded by weaker 
hydrogen bonds and provide a different class of systems to test the performance of the
LSIC method. Earlier studies using LSIC($z$) on the polarizabilities and ionization 
have shown that LSIC($z$) provides an excellent descriptions of these properties 
when compared to the CCSD(T) results\cite{C9CP06106A,waterpolarizability}.
Here, we study the binding energies of the water clusters. We find that
the choice of iso-orbital indicator plays crucial role in water
cluster binding energies.
The structures considered in this work are (H$_2$O)$_n$ ($n=1-6$) whose geometries 
are from the WATER27 set\cite{doi:10.1021/ct800549f} optimized at the B3LYP/6-311++G(2d,2p)
theory.  The hexamer structure has a few known isomers, and we considered the book (b), 
cage (c), prism (p), and ring (r) isomers. 
The results are compared against the CCSD(T)-F12b values from Ref.~[\onlinecite{doi:10.1021/acs.jctc.6b01046}] 
in Table \ref{tab:waterbinding}. 
We obtained the MAEs of  118.9, 172.1, and 46.9 meV/H$_2$O for PZSIC, LSIC($z$), and LSIC($w$), respectively.
Thus, LSIC($z$) underestimates binding energies of water cluster by roughly similar magnitude as
LSDA (MAE, 183.4 meV/H$_2$O). This is one case where LSIC($z$) does not improve over PZSIC.
A simple explanation for this behavior of LSIC($z$) is that although $z_\sigma$ used in LSIC($z$) 
can detect the weak bond regions, it ($z_\sigma$) cannot differentiate the slow-varying density regions 
from weak bond regions.  The $z_\sigma \rightarrow 0$ in the both situations causing the 
weak regions to be improperly treated. 
Fig. \ref{fig:waterbind} (a) shows $z_\sigma$ for water dimer where both slow-varying density and weak interaction regions are detected but not differentiated.
As a result, the total energies of the complex shift too much in comparison to the individual water molecules.
Thus, the underestimation of water cluster binding
energies is due to the choice of $z$ and not the LSIC method. Indeed by choosing the $w$ as a 
scaling parameter, the binding energies are much improved.
Fig. \ref{fig:waterbind} (b) shows there is no discontinuity of $w$ between 
the two water molecules ($w_i$'s of two FLOs along the hydrogen bond are plotted together in the figure). Hence 
unlike in LSIC($z$), weak interacting region is not improperly scaled down with LSIC($w$).
LSIC($w$) shows MAE of 46.9 meV/H$_2$O comparable to SCAN (MAE, 35.2 meV/H$_2$O). This result is interesting 
as  SCAN uses a function that can identify weak bond interaction. So  LSIC($w$)-LSDA may be behaving qualitatively
similar to the detection function in SCAN in weak bond regions.
The study of water binding energies is so far a unique case where the original LSIC($z$) performed poorly.
But LSIC can be improved by simply using a different iso-orbital indicator. This case serves as a motivation 
in identifying appropriate iso-orbital indicator that would work for all bonding regions in LSIC.
\begin{table}
\caption{\label{tab:waterbinding}The binding energy of water clusters (in meV/H$_2$O).}
\begin{tabular*}{0.68\textwidth}{@{\extracolsep{\fill}}lcccc}
\toprule
$n$   &   PZSIC  &   LSIC($z$)    &   LSIC($w$)   & CCSD(T)$^\textit{a}$\\ 
\midrule
2	&   -153.7	&   -34.9	&   -82.7	&   -108.6 \\
3	&   -321.6	&   -73.9	&   -183.0	&   -228.4 \\
4	&   -425.2	&   -125.0	&   -248.6	&   -297.0 \\
5	&   -446.9	&   -142.7	&   -264.8	&   -311.4 \\
6b	&   -467.1	&   -133.6	&   -275.0	&   -327.3 \\
6c	&   -466.8	&   -113.9	&   -274.8	&   -330.5 \\
6p	&   -467.7	&   -104.8	&   -276.2	&   -332.4 \\
6r	&   -458.1	&   -150.5	&   -275.5	&   -320.1 \\ 
\midrule
MAE	&   118.9	&   172.1	&   46.9    &  \\
\bottomrule
%\end{tabular*}
%\begin{flushleft}
\multicolumn{5}{l}{$^\textit{a}$Reference~[\onlinecite{doi:10.1021/acs.jctc.6b01046}]}\\
%\end{flushleft}
\end{tabular*}
\end{table}

\begin{figure}
    \centering
    \includegraphics[width=0.8\columnwidth]{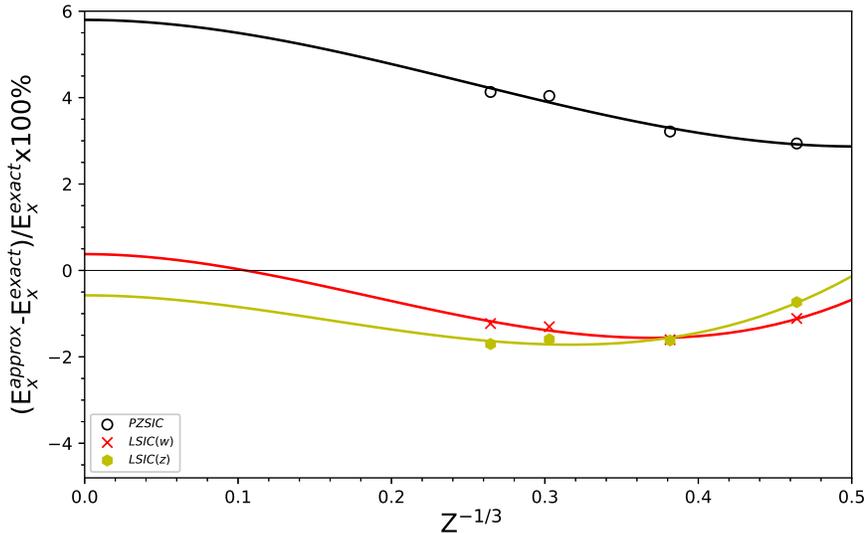}
    \caption{
    Plot of percentage error of the approximated exchange energy compared to the exact exchange energy as a function of $Z^{-1/3}$.}
    \label{fig:unifform_gas_lim}
\end{figure}

We now provide a qualitative explanation of why LSIC($w$) gives improved results over PZSIC. This reasoning 
is along the same line as offered by Zope \textit{et al.} \cite{doi:10.1063/1.5129533}.
As mentioned in Sec. \ref{sec:introduction}, when the self-interaction-errors are removed using PZSIC, improved description of barrier 
heights which involve stretch bonds is obtained but the equilibrium properties like total energies, atomization energies
etc. are usually deteriorated compared to the uncorrected functional. This is especially so for the functionals 
that go beyond  the simple LSDA. Typically this is because of over correcting tendency of PZSIC.
The non-empirical semilocal DFA functionals are designed to be exact in the uniform electron gas limit, this 
exact condition is no longer satisfied when PZSIC is  applied to the functionals\cite{doi:10.1063/1.5090534}.
This can be seen from the exchange energies of noble gas atoms and the extrapolation using the large-$Z$ expansion
of $E_X$ as shown in Fig.  \ref{fig:unifform_gas_lim}. Following Ref.~[\onlinecite{doi:10.1063/1.5090534}]
we plot atomic exchange energies  as a function of $z^{-1/3}$. Thus, the region
near origin corresponds to the uniform gas limit. The plot was obtained by fitting the exchange
exchange energies ($E_X$)  of Ne, Ar, Kr, and Xe atoms (within LSIC($w$)-LSDA, LSIC($z$)-LSDA, and LSDA) 
for Ne, Ar, Kr, and Xe atoms to the curve using the following fitting function\cite{doi:10.1063/1.5090534}.
\begin{equation}
    \frac{E_X^{approx}-E_X^{exact}}{E_X^{exact}}\times 100\%=a+bx^2+cx^3,
\end{equation}
where $x=Z^{-1/3}$ and a, b, and c are the fitting parameters. 
The LSDA is exact in the uniform gas limit. So also is LSIC($z$) since the scaling
factor $z_\sigma$ approaches zero as the gradient of electron density vanishes while 
the kinetic energy density in the denominator remains finite.
The small deviation from zero seen near origin (in Fig. \ref{fig:unifform_gas_lim})
for LSIC($z$) is due to the fitting  error (due to limited data point). 
This error is $-0.62\%$ for LSIC($z$).
Thus correcting LSDA using PZSIC introduces large error in the uniform gas limit.
The scaling factor $w$  used here identifies single-electron region since the 
density ratio approaches one in this limit. Fig.  \ref{fig:unifform_gas_lim}
shows that present LSIC($w$) approach also recovers the lost uniform limit gas.
This partly explains the success of LSIC($w$). Though performance of 
LSIC ($w$) is substantially better than PZSIC-LSDA it falls short of LSIC($z$).
On the other hand, unlike LSIC($z$) it provides good description of weak hydrogen
bonds highlighting the need of identifying suitable iso-orbital indicators or 
scaling factor(s) to apply pointwise SIC using LSIC method. One possible choice 
may be scaling factor that are functions of $\alpha$ used in construction of 
SCAN meta-GGA and recently proposed\cite{PhysRevB.99.041119} $\beta$ parameter.
A scaling factor containing $\beta$  recently used by Yamamoto and coworkers
with OSIC scheme showed improved results\cite{doi:10.1063/5.0004738}.
Future work would involve designing 
suitable scaling factors involving $\beta$ for use in LSIC method.

\section{\label{sec:conclusion} Conclusions}
To recapitulate, we investigated the performance of LSIC with a simple scaling factor, $w$, that depends
only on orbital and spin densities. Performance assessment has been carried out on 
atomic energies, atomization energies, ionization potentials, electron affinities,
barrier heights, dissociation energies etc on standard data sets of molecules.
The results show that 
LSIC($w$) performs better than PZSIC for all properties with  exception
of electron affinity and a SIE4x4 subset of dissociation energies.
We also compared the performance of $w$ for LSIC against OSIC
of Vydrov \textit{et al}.
Results indicate that although OSIC overall performs better than 
PZSIC, the improvement over PZSIC is somewhat limited.
On the other hand,  LSIC($w$) is consistently better than OSIC($w$).
We have also studied the binding energies of small water clusters which 
are bonded by weak hydrogen bonds. Here, the LSIC($w$) performs very well
compared to both the PZSIC and LSIC($z$) with performance comparable to SCAN.
The present work shows the promise of LSIC method and also demonstrates its 
limitation in describing weak hydrogen bonds if used with kinetic energy
ratio, $z_\sigma$ as an iso-orbital indicator.
This limitation is due to inability of $z_\sigma$ to distinguish weak bonding regions 
from slowly varying density regions. The scaling factor $w$ works differently
than the scaling factor $z$, hence LSIC($w$) provides good description of 
weak hydrogen bonds in water clusters.
The work thus highlights importance of designing  suitable iso-orbital indicator
for use with LSIC that can detect weak bonding regions.

\section*{Data Availability Statement}
The data that supports the findings of this study are available within the article and the supplementary information.  

\section*{Conflicts of interest}
There are no conflicts of interest to declare.

\section*{Acknowledgement}
Authors acknowledge Drs. Luis Basurto, Carlos Diaz, and Po-Hao Chang
for discussions and technical supports. 
This work was supported by the US Department of Energy, Office of 
Science, Office of Basic Energy Sciences, as part of the 
Computational Chemical Sciences Program under Award No. 
DE-SC0018331. 
Support for computational time at the Texas Advanced 
Computing Center through NSF Grant No. TG-DMR090071, 
and at NERSC is gratefully acknowledged.

%\nocite{*}
\clearpage
\bibliography{bibtex}
\bibliographystyle{rsc}

\end{document}